\documentclass[twocolumn,superscriptaddress,showpacs,preprintnumbers,amsmath,amssymb,prd]{revtex4}
\usepackage{graphicx}
\usepackage{amsmath}\usepackage{slashed}
\usepackage{color}
\usepackage{mathtools}
\usepackage{txfonts}

\UseRawInputEncoding
\begin{document}

\thispagestyle{empty}

\title{Polarization tensor in spacetime of three dimensions and quantum field
theoretical description of the nonequilibrium Casimir force in
graphene systems}

\author{
G.~L.~Klimchitskaya}
\affiliation{Central Astronomical Observatory at Pulkovo of the Russian Academy of Sciences, St.Petersburg,
196140, Russia}
\affiliation{Peter the Great Saint Petersburg
Polytechnic University, Saint Petersburg, 195251, Russia}

\author{C.~C.~Korikov}
\affiliation{Huawei Noah's Ark Lab, Krylatskaya str. 17, Moscow 121614, Russia}

\author{
V.~M.~Mostepanenko}
\affiliation{Central Astronomical Observatory at Pulkovo of the Russian Academy of Sciences, St.Petersburg,
196140, Russia}
\affiliation{Peter the Great Saint Petersburg
Polytechnic University, Saint Petersburg, 195251, Russia}

\begin{abstract}
The polarization tensor of graphene derived in the framework of the
Dirac model using the methods of thermal quantum field theory in (2+1)
dimensions is recast in a mathematically equivalent but more compact
and convenient in computations form along the real frequency axis.
The obtained unified expressions for the components of the polarization
tensor are equally applicable in the regions of the on- and
off-the-mass-shell electromagnetic waves. The advantages of the
presented formalism are demonstrated on the example of nonequilibrium
Casimir force in the configuration of two parallel graphene-coated
dielectric plates one of which is either hotter or colder than the
environment. This force is investigated as a function of temperature,
the energy gap, and chemical potential of graphene coatings with
account of the effects of spatial dispersion. Besides the
thermodynamically nonequilibrium Casimir and Casimir-Polder forces,
the obtained form of the polarization tensor can be useful for
investigation of many diverse physical phenomena in graphene
systems, such as surface plasmons, reflectances, electrical
conductivity, radiation heat transfer, etc.
\end{abstract}

\maketitle

\section{Introduction}

An investigation of the physical properties of graphene, the two-dimensional
sheet of carbon atoms, has inspired a renewed interest in the low-dimensional
quantum field theory. The point is that at energies below approximately 3~eV
graphene can be considered as a set of massless or very light quasiparticles
described by the Dirac equation where the Fermi velocity $v_F$ acts as the
speed of light $c$ \cite{1,2,3,4}. This distinguishing feature converts
graphene into a powerful tool accommodated on a laboratory table which can
be used for testing such effects of fundamental physics as the Klein
paradox \cite{5}, relativistic quantum Hall effect \cite{6}, creation
from vacuum of the particle-antiparticle pairs in external
fields \cite{7,8,9,10,11,12} etc. Further still, a strong dependence of
the dielectric properties of graphene on temperature gives no way of full
understanding the reaction of graphene to the electromagnetic field
without invoking the thermal quantum field theory in (2+1) dimensions.

The reaction of electrons or electronic quasiparticles to the
electromagnetic field is described by the one-loop polarization tensor
which has long been calculated in the frames of (2+1) dimensional
quantum electrodynamics at zero temperature\cite{13,14} (see also
Refs.~\cite{15,16,17}).
Different aspects of the polarization tensor in application to graphene,
including the case of nonzero temperature, were discussed in
Refs.~\cite{18,19,20,21}.

The reflection coefficients on a graphene sheet at zero temperature
were expressed \cite{22} via the polarization tensor of Refs.~\cite{13,14}.
The values of these reflection coefficients along the imaginary frequency
axis were used for calculation of the equilibrium zero-temperature Casimir
force in graphene systems \cite{22}. In order to investigate this force at
nonzero temperature, the polarization tensor of graphene was found at the
pure imaginary Matsubara frequencies taking into account the nonzero energy
gap (mass of quasiparticles) and the possible presence of doping (i.e.,
some foreign atoms) described by the chemical potential \cite{23}. Next,
the polarization tensor of gapped graphene was analytically continued to
the entire plane of complex frequencies including the real frequency
axis \cite{24}. A generalization of these results for the case of doped
graphene was provided in Ref.~\cite{25}. In Ref.~\cite{26}, it was shown
that the obtained tensor is unique and cannot be further modified with no
violation of the fundamental physical principles.

The expressions for the polarization tensor valid at the pure imaginary
Matsubara frequencies obtained in Ref.~\cite{23} made it possible to
calculate the thermal Casimir and Casimir-Polder interactions in many
graphene systems \cite{27,28,29,30,31,32,33}. As to the expressions for
this tensor derived in Ref.~\cite{24}, which are applicable over the
entire plane of complex frequencies, they are of multi-purpose character
and were used not only for calculation of the thermal Casimir and
Casimir-Polder forces \cite{34,35,36,37,38,39,40,41,42} but also for
investigation of the reflectivity properties \cite{43,44,45},
electrical conductivity \cite{46,47,48,49}, and the surface
plasmons \cite{50,51,52} for graphene.

The Casimir force is the physical phenomenon determined by the
electromagnetic fluctuations which occurs in the state
of thermal equilibrium when temperatures of two interacting parallel
plates are equal to each other and also equal to the temperature of the
environment. In this case, the Casimir force is described by the Lifshitz
theory \cite{53,54,55}. The condition of thermal equilibrium may be,
however, violated. This happens, for instance, when the temperature of
at least one plate is not equal to that of the environment.

As long as the local thermal equilibrium holds, the Lifshitz theory was
generalized to the situations when the standard (global) condition of
thermal equilibrium is violated. The resulting theory allowed calculation
of the nonequilibrium Casimir force between two parallel
plates \cite{56,57,58,59,60} and the Casimir-Polder force between a small
particle and a dielectric plate \cite{61,62}. These calculations demand a
knowledge of the reflection coefficients at both the Matsubara frequencies
and along the real frequency axis. In the course of further work, the
developed theory was adapted for calculation of the nonequilibrium
Casimir and Casimir-Polder forces between the arbitrarily shaped
bodies \cite{63,64,65,66,67,68,69,70}. In all these cases, however,
it was assumed that the dielectric response of all involved materials is
temperature-independent.

In Ref.~\cite{71}, the theory of the nonequilibrium Casimir force was
generalized to situations where the dielectric permittivities of
interacting bodies may depend on temperature. The developed formalism was
applied \cite{71} to the nonequilibrium Casimir force between two metallic
plates kept at different temperatures taking into account the dependence
of the relaxation parameter on temperature. Using the same formalism, the
nonequilbrium Casimir-Polder force was considered between different atoms
and a plate made of the material which undergoes the phase transition
with increasing temperature \cite{72}.

As was noted above, the dielectric properties of graphene described by the
polarization tensor strongly depend on temperature. Because of this, it is
likely that the nonequilibrium Casimir effect in graphene systems has
considerable opportunities for both theoretical and experimental
investigation. Until the present time, few investigations have been
conducted of the nonequilibrium Casimir-Polder force between nanoparticles
and the freestanding sheets of both the pristine \cite{73} and
gapped \cite{74} graphene. The cases when nanoparticles interact with
either heated or cooled plates coated with gapped \cite{75} or both gapped
and doped \cite{76} graphene were also considered using the formalism of
the polarization tensor. However, the most interesting case of the
nonequilibrium Casimir force between two plates coated with graphene
sheets was considered only in a single paper using the Kubo
formalism \cite{77}, where the energy gap was put equal to zero and the
spatial dispersion in the dielectric response of graphene was neglected.

In this paper, we obtain the more convenient analytic form for the
polarization tensor of graphene along the real frequency axis. Unlike
the previously used forms, which express the components of the
polarization tensor in the regions of the on-the-mass-shell and
off-the-mass-shell electromagnetic waves using different functions,
here we present the more unified expressions for various relationships
between the frequency and the wave vector. These expressions are
mathematically equivalent to that ones of Refs.~\cite{24,26,42,76} but
are more convenient for calculation of the physical quantities expressed
via the polarization tensor defined along the real frequency axis.

To illustrate the advantages of the suggested form of the polarization
tensor, we investigate the nonequilibrium Casimir force in the
configuration of two parallel plates coated with real graphene sheets
characterized by the nonzero energy gap and chemical potential. The
computations are made taking into account the effects of spatial
dispersion in the dielectric response of graphene coating described by
means of the polarization tensor. In the configuration considered, the
temperature of one graphene-coated plate is the same as of the environment
and of another one can be either higher or lower than that of the
environment.

According to our results, the presence of graphene coatings increases the
magnitudes of both equilibrium and nonequilibrium Casimir pressures. The
dependence of this increase on the energy gap and chemical potential of
graphene coatings is investigated. It is shown that
the magnitude of the nonequilibrium Casimir
pressure on a cooled graphene-coated plate is less than that of the
equilibrium pressure, whereas the magnitude of the nonequilibrium pressure
on a heated plate is larger than that of the equilibrium one.
For a cooled plate, the effects of nonequilibrium are larger for a smaller
energy gap, but for a heated plate they are larger for a larger energy
gap. An impact of the energy gap decreases with increasing chemical
potential. The relative error in the nonequilibrium Casimir pressure
arising due to neglect of the spatial dispersion in the dielectric
response of graphene coating is found as the function of plate
temperature, the energy gap, and chemical potential of graphene
coatings. It increases with increasing energy gap and, for a cooled
plate coated with graphene characterized by the zero chemical potential,
may reach 50\% for the temperature of 77~K. For a heated up to 500~K
graphene-coated plate, the relative error due to a neglect of the spatial
dispersion is shown to be below 9\%.

The paper is organized as follows. In Sec.~II, we present the suggested
form of the polarization tensor of graphene defined along the real frequency
axis. The previously obtained expressions at the pure imaginary Matsubara
frequencies also required in computations remain unchanged. Section III contains
the brief list of expressions for the nonequilibrium Casimir pressure
between two graphene-coated dielectric plates. In Sec.~IV, the computational
results are presented for the nonequilibrium Casimir pressure in the
configuration of two silica glass plates coated with graphene sheets
characterized by different values of the energy gap and chemical potential.
These results take full account of the spatial dispersion in graphene
coatings. Section V contains our conclusions and a discussion of the
obtained results.

Below we do not put to unity the fundamental constants $\hbar$ and $c$
in order to simplify an employment of the obtained results in various
future applications.

\section{POLARIZATION TENSOR OF GRAPHENE ALONG THE REAL FREQUENCY AXIS}

We start from the expressions for the polarization tensor of graphene
$\Pi_{\beta\gamma}(\omega,k,T)$ obtained in Refs.~\cite{24,25,26} where
$\beta,\gamma=0,1,2$,  $\omega$ is the frequency, $k$ is the magnitude of the wave
vector projection on the plane of graphene, and $T$ is the temperature of a graphene
sheet. The components of the polarization tensor also depend on the energy gap
$\Delta$ and chemical potential $\mu$ which are not indicated explicitly for
the sake of brevity.

It is convenient to express the reflection coefficients on a graphene sheet and
other quantities of physical significance via the component $\Pi_{00}(\omega,k,T)$
and the following combination of the components of the polarization tensor
\begin{equation}
\Pi(\omega,k,T)\equiv k^2\Pi_{\beta}^{\,\,\beta}(\omega,k,T)+
\left(\frac{\omega^2}{c^2}-k^2\right)\Pi_{00}(\omega,k,T).
\label{eq1}
\end{equation}

The quantities $\Pi_{00}$ and $\Pi$ are conveniently presented as the sums of two
contributions
\begin{eqnarray}
&&
\Pi_{00}(\omega,k,T)=\Pi_{00}^{(0)}(\omega,k)+\Pi_{00}^{(1)}(\omega,k,T),
\nonumber\\
&&
\Pi(\omega,k,T)=\Pi^{(0)}(\omega,k)+\Pi^{(1)}(\omega,k,T),
\label{eq2}
\end{eqnarray}
\noindent
where $\Pi_{00}^{(0)}$ and $\Pi^{(0)}$ are calculated at $T=0$, $\mu=0$.
These contributions are in fact obtained by calculating the one-loop diagram
within the standard quantum field theory at zero temperature \cite{13,14,26}.
As to the full quantities $\Pi_{00}$ and $\Pi$, they are found by calculating the same
diagram using the thermal quantum field theory in the Matsubara formulation  with
subsequent analytic continuation of the obtained results to the real frequency
axis \cite{24,26}.

We begin with the contributions $\Pi_{00}^{(0)}$ and $\Pi^{(0)}$ to Eq.~(\ref{eq2}).
In some previous literature (see, e.g., Refs.~\cite{24,26,42,76})
these contributions were
expressed differently in the regions of the off-the-mass-shell waves satisfying the
condition $\omega < v_Fk$ (the strongly evanescent waves) and $\omega \geqslant v_Fk$
(the off-the-mass-shell plasmonic waves with $v_Fk \leqslant \omega <ck$ and the
on-the-mass-shell propagating waves with $\omega \geqslant ck)$. Here we present
the following mathematically equivalent unified expressions which are valid over
the entire axis of real frequencies:
\begin{eqnarray}
&&
\Pi_{00}^{(0)}(\omega,k)=\frac{2\alpha\hbar ck^2}{\sqrt{v_F^2k^2-\omega^2}}
\Psi\left(\frac{\Delta}{\hbar\sqrt{v_F^2k^2-\omega^2}}\right),
\label{eq3}\\
&&
\Pi^{(0)}(\omega,k)=\frac{2\alpha\hbar k^2}{c}\sqrt{v_F^2k^2-\omega^2}
\,\Psi\left(\frac{\Delta}{\hbar\sqrt{v_F^2k^2-\omega^2}}\right),
\nonumber
\end{eqnarray}
\noindent
where $\alpha=e^2/(\hbar c)$ is the fine structure constant and the function $\Psi$
is defined as
\begin{equation}
\Psi(x)=x+(1-x^2)\,{\rm arctan}\left(\frac{1}{x}\right).
\label{eq4}
\end{equation}

When using Eq.~(\ref{eq3}) in different frequency regions, the branch of the square root
should be chosen as \cite{24}
\begin{equation}
\sqrt{\omega^2-v_F^2k^2}=i\sqrt{v_F^2k^2-\omega^2}.
\label{eq5}
\end{equation}
\noindent
This rule assures that the spatially nonlocal dielectric permittivities of graphene defined via
the polarization tensor have the positive imaginary parts. Note that in the frequency region
$\omega <v_Fk$ the quantities (\ref{eq3}) are real. However, if
$\omega\geqslant v_Fk$, the
quantities (\ref{eq3}) are real if the condition $\Delta >\hbar\sqrt{\omega^2-v_F^2k^2}$ is
satisfied and are complex under the opposite inequality $\Delta\leqslant\hbar\sqrt{\omega^2-v_F^2k^2}$.
The transition from real to complex values of the quantities (\ref{eq3}) corresponds to
crossing the threshold of pair creation.

Now we consider the contributions $\Pi_{00}^{(1)}$ and $\Pi^{(1)}$ to Eq.~(\ref{eq2}).
The first of them can be conveniently expressed via the following function:
\begin{equation}
X_1(x)=\frac{x^2-v_F^2k^2}{\sqrt{(\omega^2-v_F^2k^2)[x^2-v_F^2k^2A(\omega,k)]}},
\label{eq6}
\end{equation}
\noindent
where
\begin{equation}
A(\omega,k)=1-\frac{\Delta^2}{\hbar^2(\omega^2-v_F^2k^2)}.
\label{eq7}
\end{equation}

Using these notations, the quantity $\Pi_{00}^{(1)}$ in the entire region $\omega <v_Fk$
and in the region  $\omega\geqslant v_Fk$ under the condition $\hbar\sqrt{\omega^2-v_F^2k^2}<\Delta$
is given by
\begin{equation}
\Pi_{00}^{(1)}(\omega,k,T)=\frac{4\alpha\hbar c}{v_F^2}\int_{\Delta/\hbar}^{\,\infty}
\!dv\,w(v,\mu,T)
\left[1-\frac{1}{2}\sum_{\lambda=\pm 1}\lambda X_1(v+\lambda\omega)\right],
\label{eq8}
\end{equation}
\noindent
where
\begin{equation}
w(v,\mu,T)=\left[{\rm exp}\left(\frac{\hbar v+2\mu}{2k_BT}\right)+1\right]^{-1}
+\left[{\rm exp}\left(\frac{\hbar v-2\mu}{2k_BT}\right)+1\right]^{-1}.
\label{eq9}
\end{equation}

In the remaining region  $\omega\geqslant v_Fk$ under the opposite condition
$\hbar\sqrt{\omega^2-v_F^2k^2}\geqslant\Delta$ the result is
\begin{eqnarray}
&&\hspace*{-3.7mm}
\Pi_{00}^{(1)}(\omega,k,T)=\frac{4\alpha\hbar c}{v_F^2}\left\{
\int_{\Delta/\hbar}^{\,v_0}\!\!dv\,w(v,\mu,T)
\left[1-\frac{1}{2}\!\sum_{\lambda=\pm 1}\!X_1(v+\lambda\omega)\right]\right.
\nonumber \\
&&~~~
\left.
+\int_{v_0}^{\,\infty}\!dv\,w(v,\mu,T)\left[1-\frac{1}{2}\sum_{\lambda=\pm 1}\lambda
X_1(v+\lambda\omega)\right]\right\},
\label{eq10}
\end{eqnarray}
\noindent
where $v_0=\omega-v_Fk$. Note that the quantity (\ref{eq8}) considered in the region
$\omega\geqslant v_Fk$, $\hbar\sqrt{\omega^2-v_F^2k^2}<\Delta$ is real. It is, however,
complex in the entire region $\omega< v_Fk$. The quantity (\ref{eq10}) is complex as well.

The contribution $\Pi^{(1)}$ to Eq.~(\ref{eq2}) is conveniently expressed via the function
\begin{equation}
X_2(x)=\frac{\hbar^2(\omega^2-v_F^2k^2)x^2+v_F^2k^2\Delta^2}
{\hbar^2\sqrt{(\omega^2-v_F^2k^2)[x^2-v_F^2k^2A(\omega,k)]}}.
\label{eq11}
\end{equation}

As a result, in the entire region  $\omega< v_Fk$ and in the region
$\omega\geqslant v_Fk$ under
the condition $\hbar\sqrt{\omega^2-v_F^2k^2}<\Delta$, one obtains
\begin{eqnarray}
&&
\Pi^{(1)}(\omega,k,T)=\frac{4\alpha\hbar\omega^2}{cv_F^2}\int_{\Delta/\hbar}^{\,\infty}
\!dv\,w(v,\mu,T)
\nonumber\\
&&~~\times
\left[1-\frac{1}{2\omega^2}\sum_{\lambda=\pm 1}\lambda X_2(v+\lambda\omega)\right].
\label{eq12}
\end{eqnarray}
\noindent
This expression is real for $\omega\geqslant v_Fk$,
$\hbar\sqrt{\omega^2-v_F^2k^2}<\Delta$
and complex for $\omega< v_Fk$.

In the remaining region $\omega\geqslant v_Fk$,
$\hbar\sqrt{\omega^2-v_F^2k^2}\geqslant \Delta$, the
result is complex
\begin{eqnarray}
&&
\Pi^{(1)}(\omega,k,T)=\frac{4\alpha\hbar\omega^2}{cv_F^2}\left\{\int_{\Delta/\hbar}^{\,v_0}
\!dv\,w(v,\mu,T)\right.
\nonumber\\
&&~~\times
\left[1-\frac{1}{2\omega^2}\sum_{\lambda=\pm 1}X_2(v+\lambda\omega)\right]
\label{eq13} \\
&&~~~
\left.
+\int_{v_0}^{\,\infty}\!dv\,w(v,\mu,T)\left[1-\frac{1}{2\omega^2}
\sum_{\lambda=\pm 1}\lambda
X_2(v+\lambda\omega)\right]\right\}.
\nonumber
\end{eqnarray}

Equations (\ref{eq8}), (\ref{eq10}), (\ref{eq12}), and (\ref{eq13}) are mathematically equivalent
to the corresponding expressions for $\Pi_{00}^{(1)}$ and $\Pi^{(1)}$ in
Refs.~\cite{24,26,42,76}
where they are written in a more complicated form.

\section{NONEQUILIBRIUM CASIMIR PRESSURE IN THE CONFIGURATION OF TWO
GRAPHENE-COATED PLATES}
%%%%%%%%%%%%%%%%%%%%%%%%%%%%%%%%%
\newcommand{\ve}{{\varepsilon}}
\newcommand{\wk}{{(\omega,k)}}
\newcommand{\twk}{{(\omega,k,T_{1,2})}}
\newcommand{\fR}{{R_{\kappa}(u,t,T_1)}}
\newcommand{\sR}{{R_{\kappa}(u,t,T_2)}}

We consider the configuration of two parallel dielectric plates spaced at the separation $a$ made of
a material with the frequency-dependent dielectric permittivity $\varepsilon$ coated by the sheets
of graphene characterized by the energy gap $\Delta$ and chemical potential $\mu$. Let the first
graphene-coated plate have the same temperature as the environment, $T_1=T_E$. The temperature
of the second graphene-coated plate, $T_2$, can be either lower or higher than the
environmental temperature $T_E$.

In this configuration, the nonequilibrium Casimir pressure on the second plate can be presented as
the sum of two contributions \cite{59,60,71}
\begin{equation}
P_{\!\!\rm neq}(a,T_1,T_2)=P_{\rm qeq}(a,T_1,T_2)+\Delta P_{\!\!\rm neq}(a,T_1,T_2),
\label{eq14}
\end{equation}
\noindent
where $P_{\!\!\rm qeq}$ can be called the quasi equilibrium contribution and
$\Delta P_{\!\!\rm neq}$ ---
the proper nonequilibrium contribution.

A few words about the assumptions used in derivation of Eq.~(\ref{eq14}) are in order.
It has been known that the standard Lifshitz formula describing the equilibrium
Casimir force was originally derived \cite{53,54,55} using the correlations of the
polarization field expressed via the fluctuation-dissipation theorem.
These correlations are spatially local. Then, it is reasonable to admit that in the
nonequilibrium situation, where the temperatures of two plates are different,
the sources correlations are given by the same equations of the fluctuation-dissipation
theorem with the appropriate temperatures \cite{59}. This is a condition of the local
thermal equilibrium employed in the derivation of an expression for the nonequilibrium
Casimir force. Using this condition, the correlations of the electromagnetic filed
in the gap between the plates can be represented as the sum of correlations produced
by the fluctuating polarizations in the materials of the first and second plates with
the dielectric functions $\varepsilon_1$ and $\varepsilon_2$ kept at the temperatures
$T_1$ and $T_2$, respectively \cite{59}. The condition of the local thermal equilibrium
was also used in the classical paper \cite{P-VH} on the theory of radiative heat
transfer.

We begin with the proper nonequilibrium contribution which can be presented
in the form \cite{59,60,71}
\begin{widetext}
\begin{eqnarray}
&&
\Delta P_{\!\!\rm neq}=
\frac{\hbar c}{64\pi^2a^4}\int_0^{\,\infty}\! u^3du[n(u,T_1)-n(u,T_2)]
\sum_{\kappa}\left[\int_0^{\,1}\!t\sqrt{1-t^2}dt
\frac{|\sR|^2-|\fR|^2}{|D_{\kappa}(u,t,T_1,T_2)|^2}\right.
\nonumber\\
&&~~\left.
-2\int_1^{\,\infty}\!t\sqrt{t^2-1}\,e^{-u\sqrt{t^2-1}}dt\,
\frac{{\rm Im}\fR{\rm Re}\sR-
{\rm Re}\fR{\rm Im}\sR}{|D_{\kappa}(u,t,T_1,T_2)|^2}\right].
\label{eq15}
\end{eqnarray}
\end{widetext}
\noindent
Here,
\begin{eqnarray}
&&
D_{\kappa}(u,t,T_1,T_2)=1-\fR\sR\,e^{iu\sqrt{1-t^2}},
\nonumber \\
&&
n(u,T_j)=\left(\exp\frac{\hbar cu}{2ak_BT_j}-1\right)^{-1},
\quad j=1,\,2,
\label{eq16}
\end{eqnarray}
\noindent
and the dimensionless integration variables $u=2a\omega/c$
and $t=ck/\omega$
are expressed via the frequency
and the wave vector projection.

The reflection coefficients on the graphene-coated plates for the
transverse magnetic ($\kappa={\rm TM}$) and transverse electric
($\kappa={\rm TE}$) polarizations of the electromagnetic field
are given by \cite{78}
\begin{widetext}
\begin{eqnarray}
&&
R_{\rm TM}\twk=\frac{\hbar k^2[\ve(\omega)q\wk-\tilde{q}\wk]+
q\wk\tilde{q}\wk\Pi_{00}\twk}{\hbar k^2[\ve(\omega)q\wk+\tilde{q}\wk]+
q\wk\tilde{q}\wk\Pi_{00}\twk},
\nonumber\\
&&
R_{\rm TE}\twk=\frac{\hbar k^2[q\wk-\tilde{q}\wk]-
\Pi\twk}{\hbar k^2[q\wk+\tilde{q}\wk]+\Pi\twk}.
\label{eq17}
\end{eqnarray}
\end{widetext}
\noindent
In these equations, $q\wk\equiv\sqrt{k^2-\omega^2/c^2}$,
$\tilde{q}\wk\equiv\sqrt{k^2-\ve(\omega)\omega^2/c^2}$,
and the polarization tensor is defined in Eqs.~(\ref{eq2}),
(\ref{eq3}), (\ref{eq8}),  (\ref{eq10}), (\ref{eq12}), and
(\ref{eq13}). The expressions for the reflection coefficients
(\ref{eq17}) in terms of the dimensionless variables $u$
and $t$ are obtained by substituting $\omega=cu/(2a)$ and
$k=tu/(2a)$ in Eq.~(\ref{eq17}). The result is
\begin{widetext}
\begin{eqnarray}
&&
R_{\rm TM}(u,t,T_{1,2})=\frac{\hbar t^2u[\ve(u)\sqrt{t^2-1}-
\sqrt{t^2-\ve(u)}]+2a\sqrt{t^2-1}\sqrt{t^2-\ve(u)}
\Pi_{00}(u,t,T_{1,2})}{\hbar t^2u[\ve(u)\sqrt{t^2-1}+
\sqrt{t^2-\ve(u)}]+2a\sqrt{t^2-1}\sqrt{t^2-\ve(u)}
\Pi_{00}(u,t,T_{1,2})},
\nonumber\\
&&
R_{\rm TE}(u,t,T_{1,2})=\frac{\hbar t^2u^3[\sqrt{t^2-1}-
\sqrt{t^2-\ve(u)}]-8a^3\Pi(u,t,T_{1,2})}{\hbar t^2u^3[\sqrt{t^2-1}+
\sqrt{t^2-\ve(u)}]+8a^3\Pi(u,t,T_{1,2})}.
\label{eq18}
\end{eqnarray}
\end{widetext}

We continue with the quasi-equilibrium contribution $P_{\!\!\rm qeq}$ to
the nonequilibrium Casimir pressure (\ref{eq14}). This name reflects the
fact that it is calculated by the combination of the Lifshitz-type
formulas written along the imaginary frequency axis with appropriately
taken into account temperatures of the graphene-coated plates
\begin{widetext}
\begin{eqnarray}
&&
P_{\!\!\rm qeq}(a,T_1,T_2)=-\frac{k_B}{2\pi}\left[T_1
\sum_{l=0}^{\infty}{\vphantom{\sum}}^{\prime} \int_0^{\,\infty}q_{l,1}kdk
\sum_{\kappa}\frac{R_{\kappa}(i\xi_{l,1},k,T_1)
R_{\kappa}(i\xi_{l,1},k,T_2)}{e^{2q_{l,1}a}-R_{\kappa}(i\xi_{l,1},k,T_1)
R_{\kappa}(i\xi_{l,1},k,T_2)} \right.
\nonumber \\
&&~~+\left. T_2
\sum_{l=0}^{\infty}{\vphantom{\sum}}^{\prime} \int_0^{\,\infty}q_{l,2}kdk
\sum_{\kappa}\frac{R_{\kappa}(i\xi_{l,2},k,T_1)
R_{\kappa}(i\xi_{l,2},k,T_2)}{e^{2q_{l,2}a}-R_{\kappa}(i\xi_{l,2},k,T_1)
R_{\kappa}(i\xi_{l,2},k,T_2)} \right].
\label{eq19}
\end{eqnarray}
\end{widetext}
\noindent
Here, $k_B$ is the Boltzmann constant, $\xi_{l,1(2)}=2\pi k_B T_{1(2)}l/\hbar$
with $l=0,\,1,\,2,\,\ldots$ are the Matsubara frequencies calculated either
at temperature $T_1$ or $T_2$, the reflection coefficients $R_{\kappa}$
are obtained from Eq.~(\ref{eq17}) by putting $\omega=i\xi_{l,1(2)}$,
$q_{l,1(2)}\equiv q(i\xi_{l,1(2)},k)$, and the prime on the sums in $l$
divides the term with $l=0$ by 2.

The expressions for the polarization tensor at the pure imaginary frequencies
$\omega=i\xi_{l,1(2)}$ entering the reflection coefficients (\ref{eq17})
are obtained from Eq.~(\ref{eq3}) and Eqs.~(\ref{eq8}), (\ref{eq12}) valid
for the region $\omega<v_Fk$ with an appropriate choice of the branch of
square roots. We do not present them here because they are contained in many
papers (see, e.g., Refs.~\cite{36,37,39,76}).

The term quasi equilibrium regarding the quantity (\ref{eq19}) is also
justified by the fact that for the temperature-independent dielectric
response of the plate materials (this is the case, for instance, for the
silica glass plates in the absence of graphene coating) the reflection
coefficients in Eq.~({\ref{eq19}) depend on the temperature only implicitly
through the Matsubara frequencies. In this case the proper nonequilibrium
contribution (\ref{eq15}) vanishes and the quasi equilibrium contribution
(\ref{eq19}) is expressed as a half of a sum of the truly equilibrium
Casimir pressures calculated at the temperatures of the plates \cite{56}.
As a result,
\begin{equation}
P_{\!\!\rm neq}(a,T_1,T_2)=P_{\!\!\rm qeq}(a,T_1,T_2)=\frac{1}{2}\left[
P_{\!\!\rm eq}(a,T_1)+P_{\!\!\rm eq}(a,T_2)\right].
\label{eq20}
\end{equation}
\noindent
Each term in this equation is calculated by the standard Lifshitz
formula under an assumption that the environmental temperature
is the same as of the plate, i.e., $T_1$ for the first plate and
$T_2$ for the second plate.

\section{Application to nonequilibrium Casimir pressure in
configuration of two dielectric plates coated with real
graphene sheets}

As an example of the presented formalism, we consider two parallel
silica glass plates coated with real graphene sheets. The presence
of a substrate along with the influence of foreign atoms and
electron-electron interactions give rise to some nonzero energy gap
$\Delta$ in the spectrum of graphene quasiparticles \cite{20,79,80}.
As to the foreign atoms, they result in a nonzero value of the
chemical potential $\mu$ of graphene coating. The polarization
tensor of graphene considered above takes into account both these
parameters. A dependence of the polarization tensor on the magnitude
of wave vector projection $k$ reflects the effects of spatial
dispersion in the dielectric response of graphene.

Computations of the Casimir pressure on the second plate are made
for the cases when it is either heated up  to $T_2=500~$K or
cooled down to $T_2=77~$K. In so doing the temperature of the
first plate is kept the same as that of the environment
$T_1=T_E=300~$K. The proper nonequilibrium contribution
$\Delta P_{\!\!\rm neq}$ to the Casimir pressure  $P_{\!\!\rm neq}$ was
computed by Eqs.~(\ref{eq15}) and (\ref{eq18}) where the
polarization tensor of graphene coating is given by
Eqs.~(\ref{eq2}), (\ref{eq3}), (\ref{eq8}), (\ref{eq10}),
(\ref{eq12}), and (\ref{eq13}). The dielectric permittivity of
the plate along the real frequency axis was obtained from the
tabulated optical data of silica glass \cite{81} (see
Ref.~\cite{82} for more details). The quasi equilibrium
contribution $P_{\!\!\rm qeq}$ to the Casimir pressure  $P_{\!\!\rm neq}$
was computed by Eqs.~(\ref{eq19}) and (\ref{eq17}). The dielectric
permittivity of silica glass at the pure imaginary Matsubara
frequency axis was obtained from ${\rm Im}\,\ve(\omega)$ by using
the Kramers-Kronig relation \cite{82}.

Below the computational results for the nonequilibrium Casimir
pressure are compared with those for the equilibrium one at
$T_E=300~$K. The latter are computed by Eqs.~(\ref{eq19}) and
(\ref{eq20}) with $T_1=T_2=T_E=300~$K. To investigate the role
of graphene coating, the obtained results for the case of
graphene-coated plates are compared with the case of uncoated
SiO${}_2$ plates. For this purpose, one should put
$\Pi_{00}=\Pi=0$ in the reflection coefficients (\ref{eq17})
and (\ref{eq18}). Finally, we investigate the effect of spatial
dispersion in the dielectric response of graphene coating on
the  Casimir pressure. To obtain the Casimir pressure in the
spatially local approximation, i.e., with the effects of
spatial dispersion disregarded, we use expressions
(\ref{eq2}), (\ref{eq3}), (\ref{eq8}), (\ref{eq10}),
(\ref{eq12}), and (\ref{eq13}) for the polarization tensor in the
limiting case $v_Fk/\omega\to 0$.

%%%%%%%%%%%%%%__FIGURE__1_____%%%%%%%%%%%%%%%%
\begin{figure}[b]
\vspace*{-5.6cm}
\hspace*{-3cm}
\includegraphics[width=6.6in]{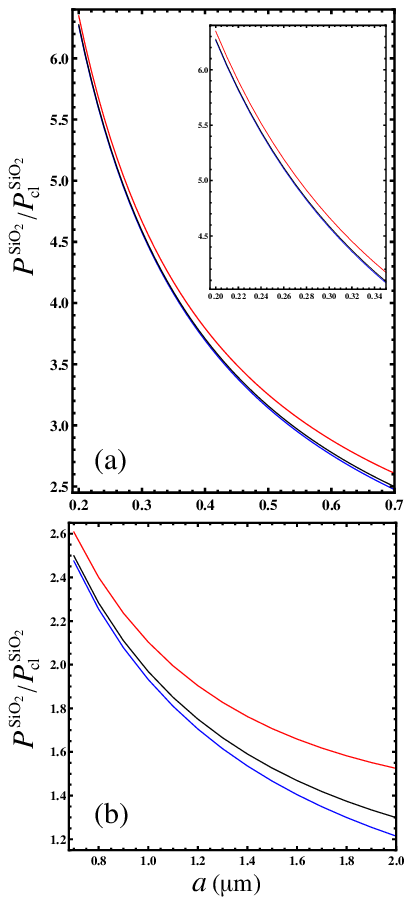}
\vspace*{-7.4cm}
\caption{\label{fg1}
The ratios of the nonequilibrium Casimir pressures for the uncoated
SiO$_2$ plates (top and bottom lines for the plate temperatures of
500~K and 77~K, respectively) and of the equilibrium pressure (the middle line)
to the classical limit of the equilibrium pressure are shown as the functions
of separation in the distance ranges (a) from 0.2 to 0.7~$\muup$m and from
0.2 to 0.35~$\muup$m in the inset on an enlarged scale and (b) from 0.7 to
2~$\muup$m. }
\end{figure}
%%%%%%%%%%%%
%%%%%%%%%%%%%%%%%%%%%%%%%%%%%%%%%%%%%%%%%%%%%%%%%%%%%%%%%%%%%%%%
In Fig.~\ref{fg1}, the role of the effects of nonequilibrium is
illustrated for the uncoated SiO${}_2$ plates. For this purpose,
the ratios of nonequilibrium pressures $P_{\!\!\rm neq}^{\,{\rm SiO}_2}$
at the temperatures of the second plate $T_2=500~$K and 77~K to
the classical limit of the equilibrium pressure at $T_E=300~$K,
$P_{\!\!\rm cl}^{\,{\rm SiO}_2}$, are shown by the top and bottom
lines, respectively, as the functions of separation between the plates.
Here, the classical limit of the equilibrium Casimir pressure is
given by \cite{82}
\begin{equation}
P_{\!\!\rm cl}^{\,{\rm SiO}_2}(a,T)=-\frac{k_BT}{8\pi a^3}\,
{\rm Li}_3\left[\left(\frac{\ve_0-1}{\ve_0+1}\right)^2\right],
\label{eq21}
\end{equation}
\noindent
where ${\rm Li}_3(z)$ is the polylogarithm function and the static
dielectric permittivity of silica glass is $\ve_0=3.81$.

The middle lines in Fig.~\ref{fg1} show the dependence on separation
for the equilibrium Casimir pressure
$P_{\!\!\rm eq}^{\,{\rm SiO}_2}/P_{\!\!\rm cl}^{\,{\rm SiO}_2}$.
All the dependences are shown in the distance ranges
(a) from 0.2 to 0.7~$\muup$m and (b) from  0.7 to 2~$\muup$m.
Besides that, the region of separation  from 0.2 to 0.35~$\muup$m
is shown in the inset to Fig.~\ref{fg1}(a) on an enlarged scale.
The minimum separation is chosen in order the characteristic
frequencies contributing to the Casimir pressure be well inside
the application energy range of the Dirac model. As is seen in
Fig.~\ref{fg1}, for a hotter SiO${}_2$ plate than the environment
the effects of nonequilibrium are larger  than for a colder one.

%%%%%%%%%%%%%%__FIGURE__2_____%%%%%%%%%%%%%%%%
\begin{figure}[b]
\vspace*{-7.5cm}
\hspace*{-3cm}
\includegraphics[width=6.in]{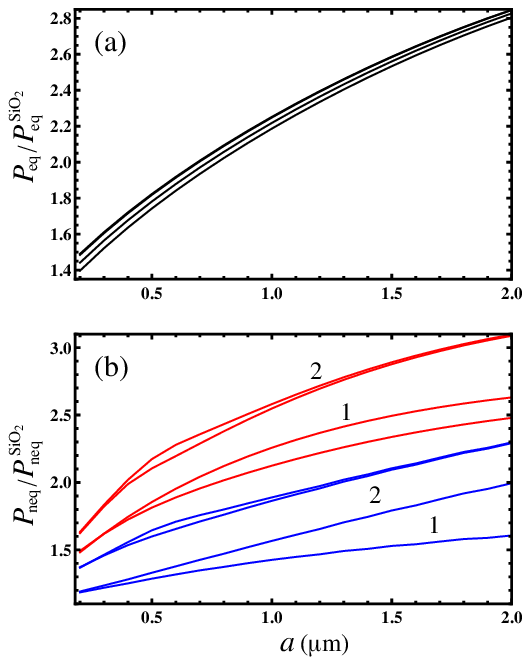}
\vspace*{-7.cm}
\caption{\label{fg2}
The ratios of the (a) equilibrium and (b) nonequilibrium Casimir
pressures for the graphene-coated SiO$_2$ plates to the (a) equilibrium and
(b) nonequilibrium Casimir pressures for the uncoated plates are shown as the
functions of separation with the following parameters of graphene coatings:
(a) the four lines counted from bottom to top are computed for $\mu=0,
\Delta=0.2$~eV; $\mu=0, \Delta=0.1$~eV; $\mu=0.25$~eV, $\Delta=0.1$~eV;
$\mu=0.25$~eV, $\Delta=0.2$~eV and (b) the bottom and top pairs of lines
labeled 1 and 2 are computed at $T=77$~K and 500~K, respectively; for the
pairs of lines 1 $\mu=0$ and for the pairs of lines 2 $\mu=0.25$~eV;
in each pair, $\Delta=0.2$ and 0.1~eV for the lower and upper lines,
respectively.  }
\end{figure}
%%%%%%%%%%%%
In Fig.~\ref{fg2}, the impact of graphene coating on the Casimir
pressure is illustrated for (a) equilibrium and (b) nonequilibrium
pressures. Thus, Fig.~\ref{fg2}(a) shows the ratio
$P_{\!\!\rm eq}/P_{\!\!\rm eq}^{\,{\rm SiO}_2}$
as the function of separation by the four lines counted from
bottom to top. The bottom line is computed for the chemical
potential $\mu=0$ and the energy gap $\Delta=0.2~$eV of graphene
coating, the middle line --- for $\mu=0$, $\Delta=0.1~$eV, and
the two overlapping top lines for $\mu=0.25~$eV, $\Delta=0.1$
and 0.2~eV. Here, $P_{\!\!\rm eq}$ is the Casimir pressure on
a graphene-coated plate and $P_{\!\!\rm eq}^{\,{\rm SiO}_2}$
on an uncoated one which is shown by the middle line in
Fig.~\ref{fg1}. It is seen that the graphene coating increases
the magnitude of the equilibrium Casimir pressure and this
increase is more marked for a smaller energy gap and larger
chemical potential.

In Fig.~\ref{fg2}(b), the ratio
$P_{\!\!\rm neq}/P_{\!\!\rm neq}^{\,{\rm SiO}_2}$ is shown
as the function of separation by the two bottom pairs of lines
1 and 2 computed at $T_2=77~$K and two top pairs of lines
1 and 2 computed at $T_2=500~$K. For the pairs of lines labeled~1,
the chemical potential $\mu=0$ and for the pairs of lines labeled~2
--- $\mu=0.25~$eV. In doing so, in each pair the lower line was
computed with the energy gap of graphene coating $\Delta=0.2~$eV
and the upper line --- with $\Delta=0.1~$eV. Note that the
values of $P_{\!\!\rm neq}^{\,{\rm SiO}_2}$ for an uncoated plate
are given by the bottom and top lines in Fig.~\ref{fg1} for
$T_2=77~$K and $T_2=500~$K, respectively. As is seen in
Fig.~\ref{fg2}(b), the magnitude of nonequilibrium Casimir
pressure is also increased by the graphene coating. This increase
is greater for higher temperature than that of the environment
and for larger chemical potential of graphene coating. Decrease
of the energy gap makes an impact of graphene coating on the
nonequilibrium Casimir pressure stronger.

We consider next the relative error arising in both equilibrium
and nonequilibrium Casimir pressures in the configuration  of two
graphene-coated plates when the dielectric response of graphene
coating is described in the spatially local approximation, i.e.,
with no regard for the spatial dispersion. In the equilibrium case,
this error is given by
\begin{equation}
\delta P_{\!\!\rm eq}^{\,\rm loc}(a,T_E)=
\frac{ P_{\!\!\rm eq}^{\,\rm loc}(a,T_E)-
P_{\!\!\rm eq}(a,T_E)}{P_{\!\!\rm eq}(a,T_E)}.
\label{eq22}
\end{equation}
\noindent
Here, in order to calculate $P_{\!\!\rm eq}^{\,\rm loc}$, the limit
of $v_Fk/\omega\to 0$ in the polarization tensor should be taken
first. The obtained expressions are considered at the pure imaginary
Matsubara frequencies $\omega=i\xi_l$ as discussed in Sec.~III.

%%%%%%%%%%%%%%__FIGURE__3_____%%%%%%%%%%%%%%%%
\begin{figure}[t]
\vspace*{-6.cm}
\hspace*{-2cm}
\includegraphics[width=5.in]{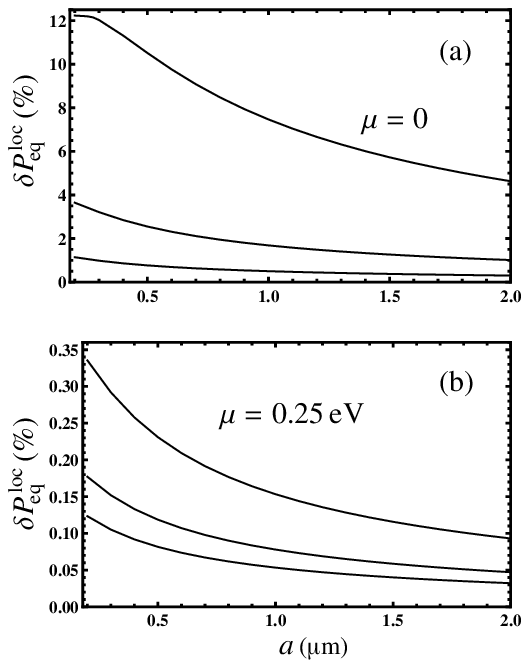}
\vspace*{-6.cm}
\caption{\label{fg3} The relative error in the equilibrium Casimir pressure for the
graphene-coated plates arising when using the spatially local approximation
in the dielectric response of graphene coatings is shown as the function
of separation by the three lines counted from bottom to top for the values
of the energy gap $\Delta=0.1$, 0.2, and 0.3~eV, respectively, and of the
chemical potential (a) $\mu=0$ and (b) $\mu=0.25$~eV.
 }
\end{figure}
%%%%%%%%%%%%
In Fig.~\ref{fg3}, the relative error (\ref{eq22}) in the equilibrium
Casimir pressure arising when using the local approximation is shown
as the function of separation by the three lines counted from bottom
to top for the energy gap of graphene coatings $\Delta=0.1$, 0.2, and
0.3~eV, respectively, whereas the chemical potential is equal to
(a) $\mu=0$ and (b) $\mu=0.25~$eV. As is seen in Fig.~\ref{fg3}(a,b),
the error due to using the local approximation in the dielectric
response of graphene in the state of thermal equilibrium increases
with increasing energy gap and decreases with increasing separation.
For the graphene coating with zero chemical potential it reaches 12\%
at the shortest separation but becomes less than a fraction of a percent
for graphene coating with $\mu=0.25~$eV.

We are coming now to the relative error in the nonequilibrium Casimir
pressure which arises from using the spatially local description of
the dielectric response of graphene coating. This error is described
by the quantity.
\begin{equation}
\delta P_{\!\!\rm neq}^{\,\rm loc}(a,T_1,T_2)=
\frac{ P_{\!\!\rm neq}^{\,\rm loc}(a,T_1,T_2)-
P_{\!\!\rm neq}(a,T_1,T_2)}{P_{\!\!\rm neq}(a,T_1,T_2)}.
\label{eq23}
\end{equation}

%%%%%%%%%%%%%%__FIGURE__4_____%%%%%%%%%%%%%%%%
\begin{figure}[b]
\vspace*{-6cm}
\hspace*{-2cm}
\includegraphics[width=5.0in]{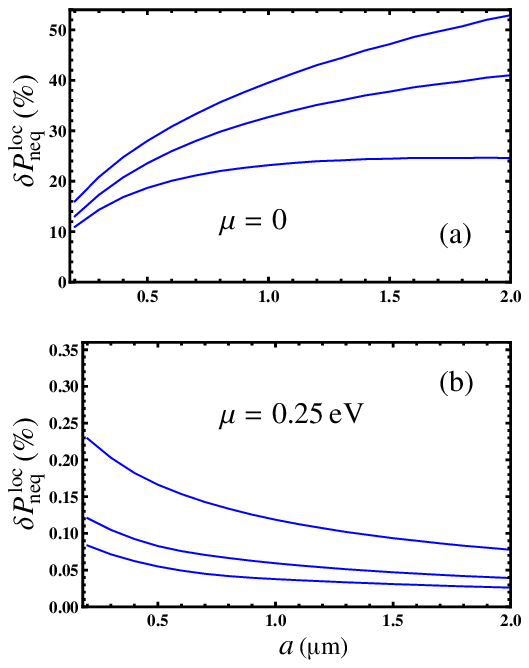}
\vspace*{-6.cm}
\caption{\label{fg4}
The relative error in the nonequilibrium Casimir pressure for the
graphene-coated plate cooled to 77~K, which arises when using the spatially
local approximation in the dielectric response of graphene coatings,
is shown as the function of separation by the three lines counted from
bottom to top for the values of the energy gap $\Delta=0.1$, 0.2, and 0.3~eV,
respectively, and of the chemical potential (a) $\mu=0$ and (b) $\mu=0.25$~eV. }
\end{figure}
%%%%%%%%%%%%
In Fig.~\ref{fg4}, the computational results for
$\delta P_{\!\!\rm neq}^{\,\rm loc}$ for the second plate cooled down
to $T_2=77~$K are presented as the functions of separation by the three
lines counted from bottom to top for the energy gap of graphene coating
$\Delta=0.1$, 0.2, and 0.3~eV, respectively, and the chemical potential
(a) $\mu=0$ and (b) $\mu=0.25~$eV (recall that $T_1=T_E=300~$K).
As is seen in Fig.~\ref{fg4}(a,b), for the cooled graphene coating with
$\mu=0$ the error arising from using the spatially local description
increases with increasing energy gap and separation within the separation
range considered. Thus, for $a=2~\muup$m and $\Delta=0.3~$eV it exceeds
50\%. At the same time, for a graphene coating with $\mu=0.25~$eV,
this error decreases with increasing separation and takes the values
below 1\% even for largest value of the energy gap considered.

%%%%%%%%%%%%%%__FIGURE__5_____%%%%%%%%%%%%%%%%
\begin{figure}[b]
\vspace*{-6cm}
\hspace*{-2cm}
\includegraphics[width=5.0in]{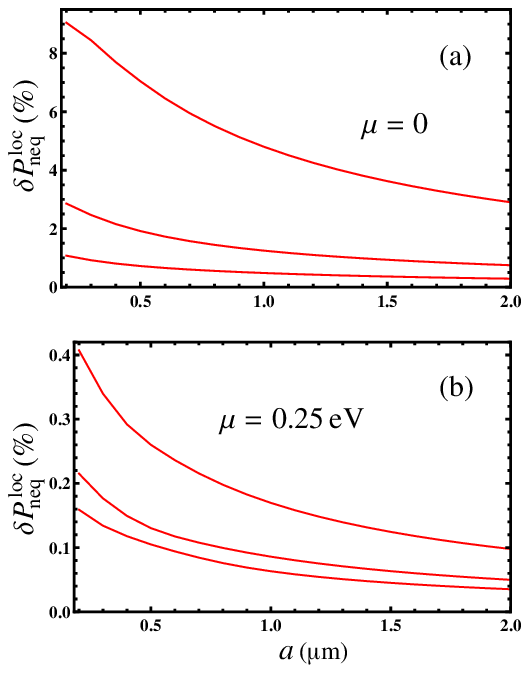}
\vspace*{-6.cm}
\caption{\label{fg5}
The relative error in the nonequilibrium Casimir pressure for the
graphene-coated plate heated to 500~K, which arises when using the spatially
local approximation in the dielectric response of graphene coatings,
is shown as the function of separation by the three lines counted from
bottom to top for the values of the energy gap $\Delta=0.1$, 0.2, and 0.3~eV,
respectively, and of the chemical potential (a) $\mu=0$ and (b) $\mu=0.25$~eV.
 }
\end{figure}
%%%%%%%%%%%%
The computational results for
$\delta P_{\!\!\rm neq}^{\,\rm loc}$ for the heated second plate up
to $T_2=500~$K are presented in Fig.~\ref{fg5} in the same form and
using the same notations as in Fig.~\ref{fg4}. From  Fig.~\ref{fg5}(a,b)
it is seen that in the case of a heated plate the error of the
spatially local approximation again increases with increasing energy
gap but decreases with increasing separation between the plates for
both values of the chemical potential considered. The maximum value
of this error of 9\% is reached for the graphene coating with
$\mu=0$ at $a=0.2~\muup$m. For the doped graphene coatings with
$\mu=0.25~$eV, the error from using the spatially local
approximation is again below a fraction of a percent.

Finally we consider the role of the effects of nonequilibrium in the
Casimir pressure on a graphene-coated plate when the exact calculation
method is used. For this purpose, we consider the ratio
$P_{\!\!\rm neq}/P_{\!\!\rm eq}$ for different values of the energy gap
and chemical potential of graphene coatings, where $P_{\!\!\rm neq}$
is calculated at either $T_2=77~$K or 500~K and $P_{\!\!\rm eq}$
at $T_1=T_2=T_E=300~$K.

%%%%%%%%%%%%%%__FIGURE__6_____%%%%%%%%%%%%%%%%
\begin{figure}[t]
\vspace*{-6.5cm}
\hspace*{-2cm}
\includegraphics[width=5.5in]{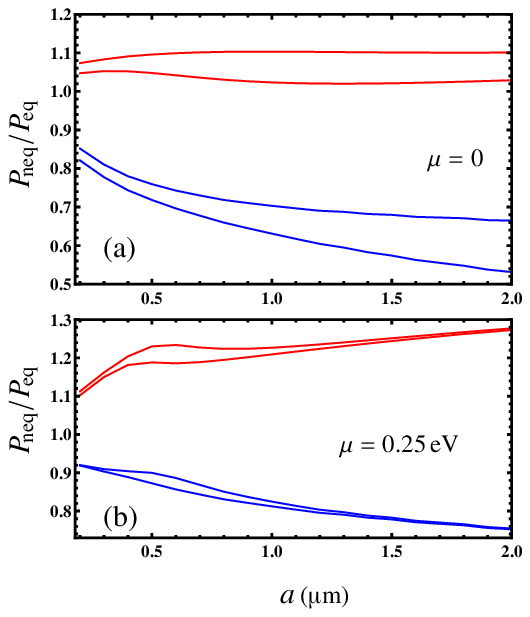}
\vspace*{-7.cm}
\caption{\label{fg6}
The ratio of the nonequilibrium Casimir pressure for the
graphene-coated SiO$_2$ plates computed using the exact theory to the
equilibrium one is shown as the function of separation by the top and bottom
pairs of lines for the plate temperatures 500~K and 77~K, respectively,
and the chemical potential (a) $\mu=0$ and (b) $\mu=0.25$~eV. In each pair,
the energy gap is equal to $\Delta=0.1$ and 0.2~eV for the lower and upper
lines, respectively. }
\end{figure}
%%%%%%%%%%%%
The computational results for the ratio
$P_{\!\!\rm neq}/P_{\!\!\rm eq}$ are presented in Fig.~\ref{fg6}
as the functions of separation by the top and bottom pairs of lines
computed at $T_2=500~$K and 77~K, respectively, (a) for $\mu=0$
and (b) for $\mu=0.25~$eV. In each pair, the lower line was computed
for the graphene coating with $\Delta=0.1~$eV and the upper line ---
with $\Delta=0.2~$eV using the exact formalism taking into account
the spatial dispersion.

{}From Fig.~\ref{fg6}(a,b) it is seen that the effects of nonequilibrium
increase the magnitude of the equilibrium Casimir pressure for a heated
plate and decrease it for a cooled plate. For a heated plate, the increase
in the magnitude of the Casimir pressure is more pronounced for the
larger energy gap. For a cooled plate, however, the decrease in the
magnitude of the Casimir pressure is greater for a smaller energy gap.
By comparing Figs.~\ref{fg6}(a) and \ref{fg6}(b), one can conclude that
the effects of nonequilibrium make the larger impact on the equilibrium
pressure for the graphene coatings with $\mu=0.25~$eV than for $\mu=0$.
It is seen also that for larger $\mu$ an impact of the energy gap on the
effects of nonequilibrium becomes smaller.

%%%%%%%%%%%%%%%%%%%%%%%%%%%%%%%%%%%%%%%%%%%%%%%%%%%%%%
\section{Conclusions and discussion}

In the foregoing, we have presented the more compact and convenient
in computations analytic form for the polarization tensor of graphene
along the real frequency axis which can be applied to theoretical
description of many diverse phenomena in graphene systems, such as
the nonequilibrium Casimir and Casimir-Polder interactions, surface
plasmons, reflectances of graphene and graphene-coated substrates,
electrical conductivity, radiation heat transfer, etc. As an example,
we calculated the nonequilibrium Casimir pressure in the configuration
of two parallel graphene-coated plates one of which is either hotter
or colder than the environment. It should be stressed that in the
framework of the Dirac model the field theoretical formalism using the
polarization tensor takes a full account of the spatial dispersion in
graphene coatings which was disregarded previously. Worthy of mention
also are the experiments on measuring the gradient of the equilibrium
Casimir force between an Au-coated sphere and a graphene-coated plate
which were found in a very good agreement with the theoretical
predictions using the formalism of the polarization
tensor \cite{78,83,84,85}.

Using the reflection coefficients on the graphene-coated silica glass
plates expressed in terms of the frequency-dependent dielectric
permittivity of the plate material and the polarization tensor, we
performed numerical computations of the nonequilibrium Casimir pressure
on a hotter and colder plates than the environment. The cases of both
equilibrium and nonequilibrium Casimir pressures in the configuration
of uncoated silica glass plates were also considered for comparison
purposes. It was shown that graphene coating increases the magnitude
of the nonequilibrium Casimir pressure. This increase is greater for
a higher temperature $T$, larger chemical potential $\mu$ and
smaller energy gap $\Delta$ of graphene coatings.

The special attention was paid to the relative error in both the
equilibrium and nonequilibrium Casimir pressures computed in the
spatially local approximation which disregards the effects of
spatial dispersion in the dielectric response of graphene coatings.
According to our results, in the equilibrium Casimir pressure
this error increases with increasing $\Delta$ and decreases with
increasing $\mu$ and separation $a$ between the plates. For a cooled
graphene-coated plate with $\mu=0$, the relative error in the
nonequilibrium Casimir pressure increases up to 50\% with increasing
$\Delta$ and $a$. However, for a graphene coating with $\mu=0.25$~eV,
this error decreases with increasing $a$ and remains below 1\% even
for the largest value of $\Delta$ considered. Different situation
was demonstrated for a nonequilibrium Casimir pressure on a heated
graphene-coated plate. Here, the relative error due to the use of
the spatially local approximation increases with increasing $\Delta$
but decreases with increasing $\mu$ and $a$. The maximum error of
about 9\% is reached for $\mu=0$ at the shortest separation
considered.

Finally, we have found the contribution of the effects of
nonequilibrium to the total pressure by computing the ratio of the
nonequilibrium to equilibrium Casimir pressures. It was shown that
the effects of nonequilibrium increase the magnitude of the
equilibrium Casimir pressure for a hotter graphene-coated plate than
the environment and decrease it for a colder plate. For a hotter plate,
the increase is larger for larger $\Delta$ of the graphene coating but
for a colder plate the decrease in the pressure magnitude is larger for
a smaller $\Delta$. The impact of the value of $\Delta$ decreases with
increasing $\mu$.

By and large, the obtained results demonstrate that the physical
phenomena determined by the electromagnetic fluctuations in graphene
systems should be described with taken into account spatial dispersion
in the dielectric response of graphene, its temperature, chemical
potential, and the energy gap. A comprehensive description of this
kind is given in the framework of thermal quantum field theory using
the formalism of the polarization tensor. This formalism can be used
in numerous applications of graphene in both fundamental physics and
nanotechnology.

%%%%%%%%%%%%%%%%%%%%%%%%%%%%%%%%
\section*{Acknowledgments}
The work of G.L.K.\ and V.M.M.~was supported by the State assignment for basic research
(project FSEG-2023-0016).

%%%%%%%%%%%%%%%%%%%%%%%%%%%%%%%%

%%%%%%%%%%%%%%%%%%%%%%%%%%%%%%%%
\end{document}